## Research Article
# A Multifeature Fusion Approach for Power System Transient Stability Assessment Using PMU Data

Yang Li,[1] Guoqing Li,[1] Zhenhao Wang,[1] Zijiao Han,[2] and Xue Bai[2]

[1]School of Electrical Engineering, Northeast Dianli University, Jilin 132012, China
[2]State Grid Liaoning Electric Power Supply Co. Ltd., Shenyang 110006, China

Correspondence should be addressed to Yang Li; 876464104@qq.com

Taking full advantage of synchrophasors provided by GPS-based wide-area measurement system (WAMS), a novel VBpMKL-based transient stability assessment (TSA) method through multifeature fusion is proposed in this paper. First, a group of classification features reflecting the transient stability characteristics of power systems are extracted from synchrophasors, and according to the different stages of the disturbance process they are broken into three nonoverlapped subsets; then a VBpMKL-based TSA model is built using multifeature fusion through combining feature spaces corresponding to each feature subset; and finally application of the proposed model to the IEEE 39-bus system and a real-world power system is demonstrated. The novelty of the proposed approach is that it improves the classification accuracy and reliability of TSA using multifeature fusion with synchrophasors. The application results on the test systems verify the effectiveness of the proposal.

## 1. Introduction

Transient stability assessment (TSA) of a power system has always been considered as a very important work to guarantee the safe and stable operation of the system [1]. Transient stability refers to the stability of the system to keep all generators synchronous and transit to a new or restore to the original stable operation status when a large disturbance occurs [2]. Although large amounts of advanced equipment and control measures have been applied in modern power systems, blackout accidents still occur from time to time around the world [3]. Moreover, with problems arising from the interconnection operation of power grids, the large-scale integration of renewable energy and power markets, transient stability characteristics of power systems are becoming more and more complicated, and the consequences resulting from instability are growing increasingly serious therewith [1–3]. Therefore, the need for fast and accurate TSA becomes more critical, as power systems are increasingly closer to their operation limit.

Traditionally, TSA approaches can be divided into two major classes: time-domain simulation methods [4] and direct methods [5]. Time-domain simulation methods are able to provide details of the dynamic behaviors for large-scale systems. However, the assessment result of this method largely depends on the accuracy of the used system model and parameters. In addition, the high computational burden limits its online applicability. Direct methods, such as transient energy function method and extended equal area criteria, have a relatively fast computational speed and can provide quantitative indices for stability assessment [5]. Unfortunately, the approach's assessment result may not give sufficient details because of simplification of the system models. Recent research shows that single machine equivalent (SIME) transient stability methods [6, 7] and pattern recognition-based TSA (PRTSA) methods, such as decision trees (DT), artificial neural networks (ANN), and support vector machines (SVM), are promising for online TSA of power systems [8–16].

There is no doubt that postfault state information for a wide-area power system plays a very important role in TSA [17]. However, the traditional measurement systems, such as SCADA, are unable to provide wide-area synchronized phasor measurements (synchrophasors) [1, 3]. With the advent of emerging wide-area measurement system (WAMS) using global positioning system (GPS) techniques, the bottleneck is broken through. The matured applications of WAMS have made it become reality to obtain the real-time synchrophasors [18], and this brings new ideas and opportunities for implementing an advanced wide-area protection and control (WAPaC) system [19, 20]. Therefore, the present approach focuses on in-depth use and data mining from postfault dynamic information provided by WAMS.

Information fusion proves to be an effective way to improve the generalization ability and reliability of PRTSA [21–23]. However, most of the reported fusion approaches, such as classifiers ensemble, need relatively more CPU time and computing resources. A multifeature fusion approach called VBpMKL was recently proposed to solve pattern classification problems in the field of engineering [24]. VBpMKL informatively combines available multiple feature space by combining kernels, which can match and even outperform the classifier ensemble approaches while obviously reducing the computing overheads [1, 24]. In addition, the input features of the used fusion approach VBpMKL are made up of both the prefault static features and the postfault dynamic features, comprehensively indicating the transient stability of power systems during different stages of the disturbance process.

In this paper, a novel VBpMKL-based PRTSA method uses the postdisturbance PMU data to predict the system transient stability status. Different from the aforementioned SIME method in [5–7], a stability margin is calculated for a given contingency and an optimal power flow-based preventive control is then taken as countermeasures if the margin is unacceptable. The proposed approach is to trigger postfault emergency control if using PMU measurements without using any given contingencies. The contribution of this paper can be divided into two aspects: on the one hand, VBpMKL, a new multifeature fusion based on Bayesian multiple kernels learning, application to TSA using synchrophasors is demonstrated in detail; on the other hand, significant performance improvements from applying the present approach to test systems are further demonstrated.

The rest of this paper is arranged as follows. In Section 2, the basic principles of WAMS are introduced. In Section 3, the VBpMKL algorithm is briefly described. Details of the present approach using multifeature fusion with synchrophasors are shown in Section 4. In Section 5, the proposal is examined using two test systems and finally the conclusions are drawn in Section 6.

## 2. Wide-Area Measurement System

*2.1. Structure of WAMS.* WAMS is a GPS-based wide-area state monitoring system for power systems, which is based on the modern communication technology and the clock synchronization technology [17]. It is mainly made up of three parts: PMUs in a power plant or substation, a communication system covering the entire power network, and a control system in the dispatching center. The structure of diagram of WAMS is shown in Figure 1.

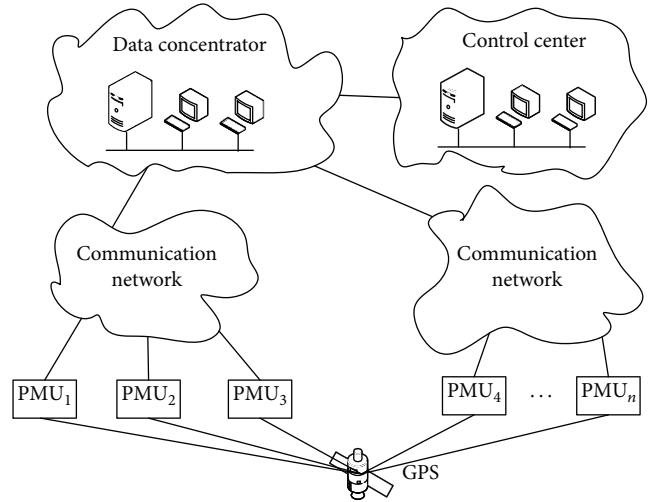

Figure 1: Structure diagram of WAMS.

The general working process of WAMS is as follows. First, through high-speed communication networks, phasor data with GPS time stamp are packaged and sent to the data concentrator by PMU substations; next, phasor data are synchronously processed and stored in the data concentrator, and then the related important parameters can be obtained, comprising synchronous voltage/current phasors, the active and reactive powers, the power factor ($\cos \varphi$), and the system frequency; finally, dynamic information of power systems can be collected quickly and effectively. In this way, WAMS is able to provide important data sources for power system monitoring, accident analysis, and stability control.

*2.2. Principle of PMUs.* As is well known, PMU is a core equipment for WAMS. It is able to measure synchronized phasor measurements from a wide-area power system with synchronized high-resolution time stamps. The hardware structure of PMU is illustrated as Figure 2.

As shown in Figure 2, the working process of PMU is as follows [18]. (1) standard time signals from the GPS receiver module are sent into the sync-signal generator and CPU module as the standard time source for data collection and phasor calculation and then sampling pulse signals from the sync-signal generator are sent to the A/D sampling module; (2) three-phase voltage/current signals from AC input are sent to CPU to compute synchrophasors through the voltage/current transformer module, the low-pass filter module, and the A/D sampling module; (3) synchrophasors are sent to the phasor data concentrator (PDC) or the control center through the commutation module.

*2.3. Application of PMU Data to PRTSA.* It is known that measured system data is crucial to determine the transient stability status of a postfault power system. By using PMU data, it became a reality to monitor and analyze the system-wide dynamic real-time behaviors of a distributed power system at the same time coordinates. Therefore, the presented

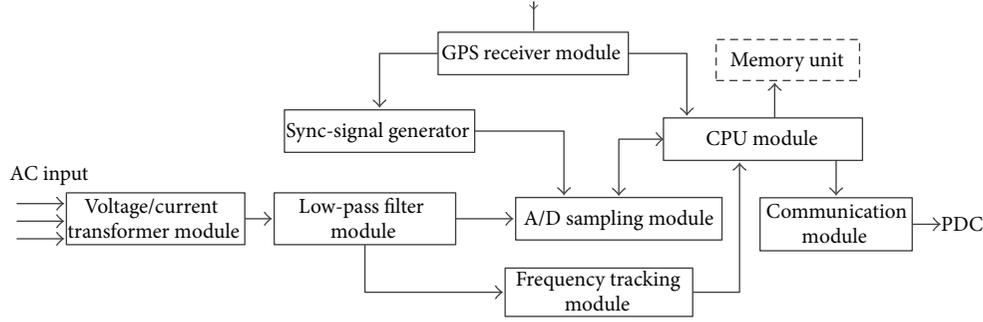

Figure 2: Hardware structure of WAMS.

approach is focused on finding new uses of PMU data to PRTSA.

In this work, based on PMU data, a set of system-level features for stability classification are extracted as predictors for PRTSA, and a VBpMKL-based TSA model is built to train the relationship between predictors and power system stability status. Here, it is supposed that tripping signal issued by the local protection is available for triggering the TSA system. Once a fault is cleared by the action of relevant relays, the trigger allows starting of taking the samples of the input variables to construct the input vector for the proposed TSA model. And then, several consecutive postfault state information of each generator from PMUs are employed to form the input vector for the model. Finally, for a specific unseen case, the transient stability status of the disturbed system can be immediately predicted by using the trained model.

Note that all the employed input features can be obtained from the following physical quantities, comprising the rotor angle, angular velocity, mechanical power, and electromagnetic power of each generator and the generators' inertial time constants; and all these physical quantities are available from PMU data except for the pregiven inertial time constants. Therefore, the proposal is able to be applied to PRTSA by using PMU data.

## 3. Introduction to VBpMKL

An object $n$ belonging to a dataset $\{\mathbf{X}, \mathbf{t}\}$, with $\mathbf{X}$ being the collection of all objects and $\mathbf{t}$ being the corresponding labels, is represented by feature spaces $S$, $D^S$-dimensional feature vectors for $\mathbf{x}_n^s$ for $s = 1, \ldots, S$ and $\mathbf{x}_n^s \in R^{D^s}$. Let the number of classes be $C$ with the target variable $t_n = c = 1, \ldots, C$ and the number of objects $N$ with $n = 1, \ldots, N$. Then, the class posterior for object $n$ will be $P(t_n \mid \mathbf{x}_n^1, \ldots, \mathbf{x}_n^S)$ and the aim of VBpMKL is to make class prediction $\mathbf{t}^*$ for $N_{\text{test}}$ new objects $\mathbf{X}^*$ that are represented by $S$.

### 3.1. Kernel Combination. Here, the $N \times N$ mean composite kernel is presented as the example

$$K^{\beta\Theta}(\mathbf{x}_i, \mathbf{x}_j) = \sum_{s=1}^{S} \beta_s K^{s\theta_s}(\mathbf{x}_i^s, \mathbf{x}_j^s) \quad \text{with} \quad \sum_{s=1}^{S} \beta_s = 1 \ \forall s, \quad (1)$$

where $\beta_s \geq 0$, $\Theta$ are the kernel parameters, $\boldsymbol{\beta}$ are the combinatorial weights, and $K$ is the kernel function employed.

In this work, Gaussian kernels and polynomial kernels are employed as base kernels corresponding to each feature space.

### 3.2. The Multinomial Probit Model. Now, we consider a linear regression model with parameters $\mathbf{W} \in R^{C \times N}$, $\mathbf{W} = [\mathbf{w}_1, \ldots, \mathbf{w}_c]^T$, and the $n$th column of the composite kernel $\mathbf{k}_n^{\beta\Theta}$ as an $N \times 1$ column vector. The multinomial probit likelihood is given by

$$P\left(t_n = i \mid \mathbf{W}, \mathbf{k}_n^{\beta\Theta}\right)$$
$$= \varepsilon_{p(u)} \left\{ \prod_{j \neq i} \Phi\left(u + \left(\mathbf{w}_i - \mathbf{w}_j\right) \mathbf{k}_n^{\beta\Theta}\right) \right\}, \quad (2)$$

where $\varepsilon$ is the expectation taken with respect to the standardized normal distribution $p(u) = N(0, 1)$ and $\Phi$ is the cumulative density function.

Using approximate Bayesian inference [1, 24], the predictive distribution for a single new object $\mathbf{x}^*$ is given by

$$p\left(t^* = c \mid \mathbf{x}^*, \mathbf{X}, \mathbf{t}\right)$$
$$= E_{p(u)} \left\{ \prod_{j \neq c} \Phi\left[\frac{1}{\overline{v_j^*}} \left(u\overline{v_c^*} + \overline{m_c^*} - \overline{m_j^*}\right)\right] \right\}. \quad (3)$$

## 4. VBpMKL-Based TSA Method

As is known, PRTSA is usually treated as a two-pattern classification problem [1, 5, 15], which includes a training phase and a test phase. It is noted that convex linear combination is chosen for the kernel combination rule in this paper. The reason for this is that, comparing with other alternatives, it usually gives the best results based on lots of experimental statistics.

### 4.1. Training Phase. The steps in training phase can be largely summarized as follows.

*Step 1.* Data preprocessing is performed using $Z$-score standardization.

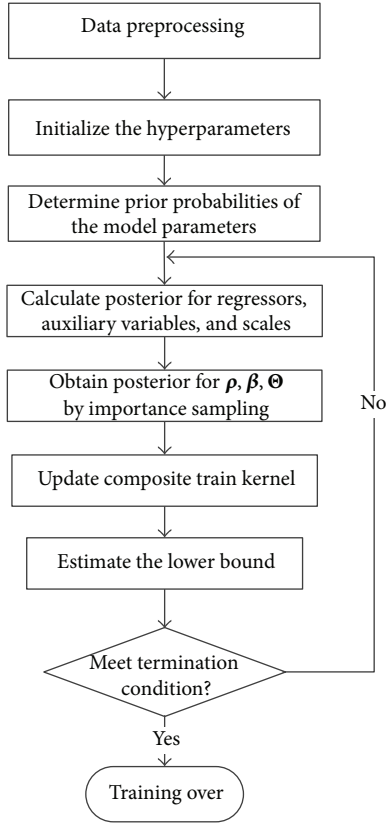

Figure 3: A flowchart of the training process.

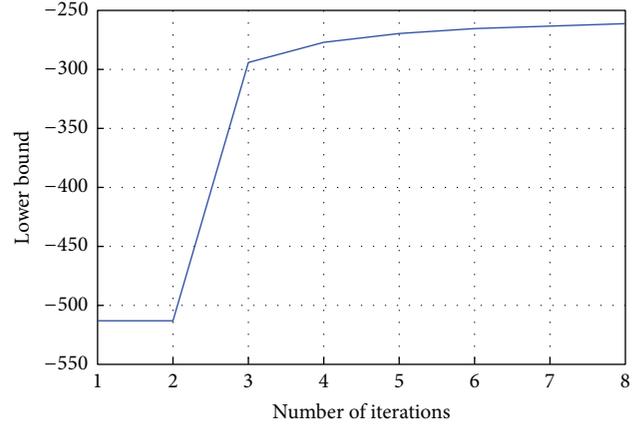

Figure 4: Change of lower bound in the training process.

*Step 2.* Initialize the hyperparameters.

*Step 3.* Determine the prior probabilities of the model parameters.

*Step 4.* Calculate posterior for regressors, auxiliary variables, and scales. According to the theory of Bayesian multiple kernels learning, the posterior for the parameters can be calculated.

*Step 5.* Obtain posterior for $\rho$, $\beta$, $\Theta$ by importance sampling, where $\rho$ is the parameters of Dirichlet distribution.

*Step 6.* Update the composite train kernel.

*Step 7.* Estimate the lower bound.

*Step 8.* Judge whether the termination condition is met or not. If the condition is met, then the training is over; else, go to Step 4.

*Step 9.* The trained TSA model is obtained.

The training process can be shown in Figure 3.
The change of lower bound in the training process is shown in Figure 4.

*4.2. Test Phase.* The steps in the test phase can be generally summarized as follows.

*Step 1.* The preprocessing approach is the same as the one used in the training phase.

*Step 2* (create composite test kernel). Creation of composite test kernel is based on the trained parameters, which are obtained after having trained the TSA model.

*Step 3* (calculate the predictive distribution). The predictive distribution for a new test sample is calculated according to (3).

*Step 4* (classify). For a given test sample, the predicted stability result is the category corresponding to the maximum value of its class membership probabilities obtained in Step 3.

## 5. Feature Selection

As we know, feature selection plays a major part in PRTSA [25–27]. The transient stability status of a postfault power system is dependent not only on the initial operating state of the prefault system but also on the disturbance's severity [2]. Thus, postfault rotor variables are suitable to be employed as predictors for PRTSA. However, the used features in previous works are mainly prefault static features. The reason for this is that the traditional measurement systems are not able to provide wide-area dynamic information. To take full advantage of synchrophasors provided by WAMS [17, 18], the presented approach selects input features from both prefault static information and postfault dynamic information.

As an extension of previous related works, the selected initial features used in the presented approach are the same as the ones in [1] (see [1] for further details). The used input features are made up of three nonoverlapped subsets: the prefault static features and the features immediately following a fault ($F_1$); the features at the fault clearing time ($F_2$); the features after the fault clearing time ($F_3$). For convenience of describing the application of the proposal in the following, the subsets are listed as follows.

*5.1. The Features of $F_1$.* The features of $F_1$ are shown in Table 1. Here, $t_0$ denotes the fault occurrence time.

Table 1: The features of $F_1$.

| Feature number | Input feature |
| --- | --- |
| Tz1 | The average mechanical power of all generators before the fault occurs |
| Tz2 | The average initial acceleration rate of all generators at $t_0$ |
| Tz3 | The mean square error of all the initial acceleration rates at $t_0$ |
| Tz4 | Average initial acceleration power of all generators at $t_0$ |
| Tz5 | The maximal initial kinetic energies of all generators at $t_0$ |
| Tz6 | The maximal initial active power impact at $t_0$ |
| Tz7 | The rotor angle of the generator which has the maximum acceleration rate at $t_0$ |

Table 2: The features of $F_2$.

| Feature number | Input feature |
| --- | --- |
| Tz8 | The system impact at $t_{cl}$ |
| Tz9 | The maximal difference of acceleration rates at $t_{cl}$ |
| Tz10 | The average kinetic energies of all generators at $t_{cl}$ |
| Tz11 | The rotor angle of the generator which has the maximum energy at $t_{cl}$ |
| Tz12 | The kinetic energy of the generator which has the maximum rotor angle at $t_{cl}$ |
| Tz13 | The maximal rotor kinetic energies at $t_{cl}$ |
| Tz14 | The total system "energy adjustment" |

In Table 1, Tz1 indicates the system static stability level in general; Tz2 represents the tendency of asynchronous operation of the disturbed power system at $t_0$; Tz3 indicates the relative stability degree among all the machines at $t_0$; Tz4 indicates the average level of supply and demand imbalances of all generators at $t_0$; Tz5 indicates the instability possibility of the most leading generator at $t_0$; Tz6 represents the destruction degree caused by the disturbance to the system at $t_0$; Tz7 indicates the quiescent operating point of the worst generator.

5.2. The Features of $F_2$. Table 2 shows the features of $F_2$, where $t_{cl}$ denotes the fault clearing time.

In Table 2, Tz8 indicates the destruction degree caused by the disturbance to the system at $t_{cl}$; Tz9 indicates the instability possibility of the worst machine at $t_{cl}$; Tz10 represents the average energy accumulation of the system at $t_{cl}$; Tz11 indicates the deceleration capability of the generator which has the maximum kinetic energy at $t_{cl}$; Tz12 indicates the instability tendency of the most leading generator at $t_{cl}$; Tz13 represents the instability possibility of the most leading generator at $t_{cl}$; Tz14 indicates the severity of the disturbance in general at $t_{cl}$.

5.3. The Features of $F_3$. The features of $F_3$ are listed as shown in Table 3, where $t_{cl+3c}$, $t_{cl+6c}$, and $t_{cl+9c}$, in turn, denote the 3rd, 6th, and 9th cycle after the fault clearance.

Table 3: The features of $F_3$.

| Feature number | Input feature |
| --- | --- |
| Tz15 | The maximal kinetic energies of all generators at $t_{cl+3c}$ |
| Tz16 | The maximal kinetic energies of all generators at $t_{cl+6c}$ |
| Tz17 | The maximal kinetic energies of all generators at $t_{cl+9c}$ |
| Tz18 | The kinetic energy of the generator which has the maximum rotor angle at $t_{cl+3c}$ |
| Tz19 | The kinetic energy of the generator which has the maximum rotor angle at $t_{cl+6c}$ |
| Tz20 | The kinetic energy of the generator which has the maximum rotor angle at $t_{cl+9c}$ |
| Tz21 | The maximal difference of all rotor angles at $t_{cl+3c}$ |
| Tz22 | The maximal difference of all rotor angles at $t_{cl+6c}$ |
| Tz23 | The maximal difference of all rotor angles at $t_{cl+9c}$ |

In Table 3, Tz15–Tz23 are the features extracted from synchrophasors, which characterize the dynamic behaviour and stability of postfault operation state of power systems.

As can be seen, the above chosen feathers ($F_1$, $F_2$, and $F_3$) comprehensively indicate the stability of power systems during different stages of the disturbance process. Therefore, they are suitably chosen as the original features for constructing the transient disturbed pattern space.

## 6. Case Study

In this section, the effectiveness of the proposal is examined using two testing cases: the IEEE 39-bus system and a large-scale real-world power system. It should be noted that synchrophasors are simulated through the detailed time-domain simulations in this work. It is also assumed that PMUs are placed on all generator buses in the two test systems. In addition, all the simulations of different algorithms are implemented in MATLAB 2008B environment, and the used simulation platform is an ordinary PC with 2.66 GHz CPU.

6.1. Case 1—The IEEE 39-Bus System. First, the system is used to test the proposal's effectiveness. The system is a widely used testing case for examining the performance of TSA approaches [1, 8, 9, 15, 16], and its single-line diagram is demonstrated in Figure 5.

6.1.1. Generation of Knowledge Base. As is known, for PRTSA, the generalization ability of a TSA model is mainly dependent on the completeness and representativeness of a knowledge base (KB). Consequently, a great quantity of detailed time-domain simulations have been implemented to create KB. The simulation calculation conditions of the modeled system are set as follows. The generator model employed is a classical machine model, and the used load model is the constant impedance model. The considered contingencies are three-phase short-circuit faults; the fault clearing time is supposed to be 5 cycles for all of the contingencies. The fault is cleared with successful reclosure and network topology is unchanged. The contingencies are repeatedly performed at 10 levels (85%, 90%, 95%, ..., 130% of the base load levels), and 5 kinds of active and reactive generator powers under each load level are randomly assigned. In addition, a total of 24 different fault locations are taken into account. Finally, a KB with total 1200 samples is created.

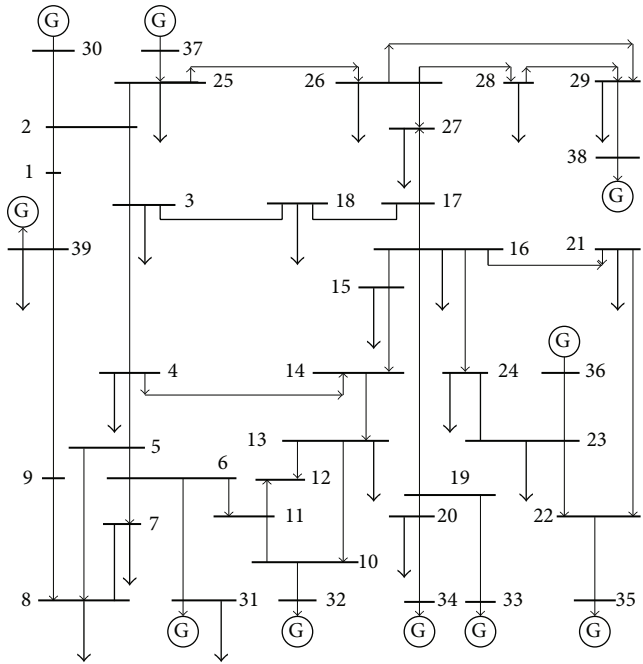

Figure 5: One-line diagram of the IEEE 39-bus system.

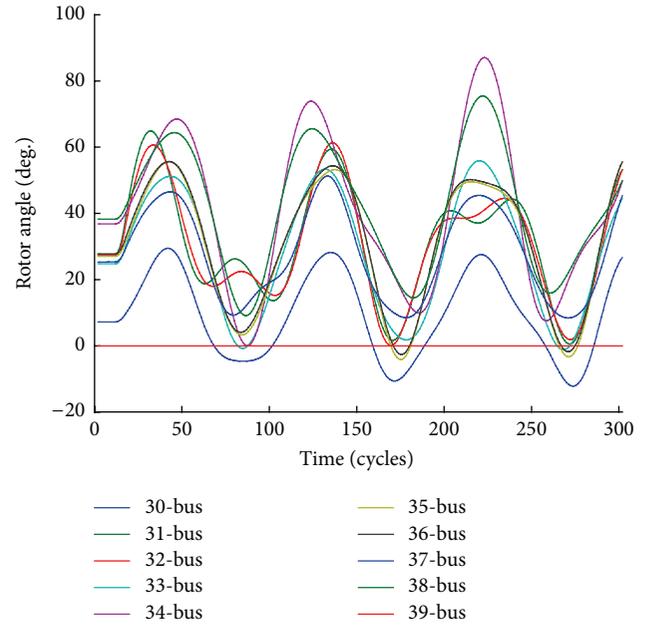

Figure 6: A transient stable case.

A class label *Class_Lable* of each sample is denoted by a transient stability index which is related to the difference of relative rotor angles in the transient process of the disturbed power system [1, 16]. The label *Class_Lable* of a sample is determined as

$$Class\_Lable = \text{sgn}\left(360° - |\Delta\delta|_{\max}\right), \quad (4)$$

where sgn(·) is a sign function, | · | is the absolute value function, and $|\Delta\delta|_{\max}$ is of the maximal difference of relative rotor angles between generators in the process. If $|\Delta\delta|_{\max} > 360°$, the *Class_Lable* is set to "−1", which represents that the system is unstable; otherwise, the label is set to "+1", which represents that the system is stable [1, 9, 16]. By plotting the swing curves of all generators, stable case and unstable cases are demonstrated in Figures 6 and 7, respectively.

Here, two groups of tests were designed and executed on two sample sets to examine the proposal reasonably. Both tests were used to test the predictive performance with different multifeature fusion schemes. Note that *Test accuracy 1* and *Test accuracy 2* in Tables 4–7 are, respectively, denoted as the test accuracy of *Test 1* and *Test 2*.

*Test 1*: from the created whole KB, 600 samples are randomly chosen as training set 1 and the rest are used as testing set 1.

*Test 2*: 900 ones out of the same KB as in *Test 1* are selected randomly as training set 2 and the rest are testing set 2.

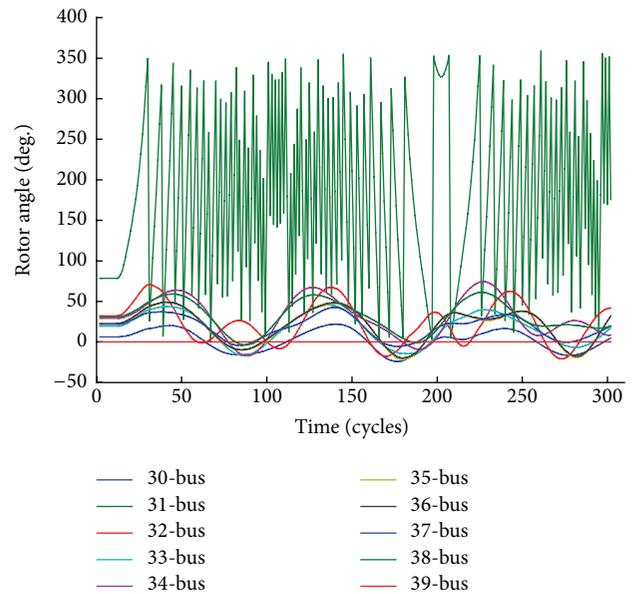

Figure 7: A transient unstable case.

*6.1.2. Effect of Different Combinations of Feature Subsets.* To evaluate the performance of different combinations of feature subsets, the present approach is tested under 8 testing schemes (Schemes 1–8) with different combinations of input feature subsets with the test results shown in Table 4. Here, Schemes 1–3 employ $F_1$, $F_2$, and $F_3$ as the data source successively, Scheme 4 employs the union of all input features as the data source, and Schemes 5–8 employ $F_1$ and $F_2$, $F_1$ and $F_3$, $F_2$ and $F_3$, and $F_1$, $F_2$, and $F_3$ as data sources, respectively. Note that all the kernel functions employed in these 8 schemes are Gaussian kernel functions.

Table 4: Test results of different combinations of feature subsets.

| Scheme number | Combination scheme | Test accuracy 1/% | Test accuracy 2/% |
|---|---|---|---|
| 1 | $F_1$ | 89.50 | 91.00 |
| 2 | $F_2$ | 90.83 | 92.67 |
| 3 | $F_3$ | 92.67 | 93.33 |
| 4 | All features | 91.33 | 93.67 |
| 5 | $F_1$ and $F_2$ | 93.67 | 94.67 |
| 6 | $F_1$ and $F_3$ | 94.50 | 96.33 |
| 7 | $F_2$ and $F_3$ | 95.50 | 97.33 |
| **8** | **$F_1$, $F_2$, and $F_3$** | **96.33** | **98.33** |

Table 5: Test results of different combinations of kernels.

| Scheme number | Combinations of kernels | Test accuracy 1/% | Test accuracy 2/% |
|---|---|---|---|
| 9 | $F_1(K_p), F_2(K_p), F_3(K_p)$ | 90.50 | 96.33 |
| 10 | $F_1(K_p), F_2(K_p), F_3(K_g)$ | 91.33 | 93.67 |
| 11 | $F_1(K_p), F_2(K_g), F_3(K_p)$ | 93.17 | 92.67 |
| 12 | $F_1(K_p), F_2(K_g), F_3(K_g)$ | 95.00 | 97.33 |
| 13 | $F_1(K_g), F_2(K_p), F_3(K_p)$ | 90.50 | 96.33 |
| 14 | $F_1(K_g), F_2(K_p), F_3(K_g)$ | 90.83 | 95.33 |
| 15 | $F_1(K_g), F_2(K_g), F_3(K_p)$ | 94.17 | 95.33 |
| **16** | **$F_1(K_g), F_2(K_g), F_3(K_g)$** | **96.33** | **98.33** |

Table 6: Test results with noisy inputs.

| Scheme number | Combinations of kernels | Test accuracy 1/% | Test accuracy 2/% |
|---|---|---|---|
| 17 | $F_1(K_g), F_2(K_g), F_3(K_g)$ | 73.67 | 76.00 |
| 18 | $F_1(K_g), F_2(K_g), F_3(K_g)$ | 95.50 | 96.67 |

Table 7: Test results of other models.

| Input features | TSA model | Test accuracy 1/% | Test accuracy 2/% |
|---|---|---|---|
| All features | DT | 93.17 | 95.33 |
| | MLP | 91.33 | 93.67 |
| | SVM | 94.50 | 96.33 |

According to the results in Table 4, it can clearly be observed that due to different combinations of feature subsets the test results of different schemes are obviously different from each other. Furthermore, the prediction performance of Scheme 8 is obviously superior to that of the other schemes. Specifically speaking, in *Test 1* and *Test 2*, the classification accuracy in Scheme 1 is lowest, which, respectively, is 89.50% and 91.00%; the one in Scheme 8 is highest, respectively, 96.33% and 98.33%; and the one in Scheme 4 is 91.33% and 93.67% in turn.

The results demonstrate that combination schemes play an important role in the final classification ability and predictive performance for the proposed VBpMKL-based TSA model. Because the input features of the feature subsets ($F_1$, $F_2$, and $F_3$) are chosen only through a preliminary analysis, all test results in Schemes 1–3 are not satisfactory. When using the union of all input features as the data source in Scheme 4, the result has still not been significantly improved. But, when using multifeature fusion through combining feature spaces, the classification accuracy in Schemes 5–8 has been evidently strengthened. Therefore, the conclusion can be drawn that multifeature fusion is an effective way to improve classification accuracy for the present approach.

*6.1.3. Effect of Different Combinations of Kernel Functions.* To test the effects of different combinations of kernel functions, 8 testing schemes (Schemes 9–16) are carried out. All these schemes employ the same combination scheme as Scheme 8 (combination of $F_1$, $F_2$, and $F_3$), and the kernel function corresponding to each feature subset, respectively, employs polynomial function ($K_p$) and Gaussian kernel function ($K_g$). The test results of these schemes are illustrated in Table 5. Please note that $F_i(K_x)$ indicates that the employed kernel function of $F_i$ is $K_x$, where $i \in \{1, 2, 3\}$ and $K_x \in \{K_p, K_g\}$.

The results in Table 5 indicate that the classification accuracy is deeply influenced by different combinations of kernel functions as well. On the one hand, the classification accuracy reaches their maximum when all three feature subsets adopt Gaussian kernel function as base kernels (Scheme 16). In this case, they are 96.33% and 98.33%, respectively. On the other hand, different combinations of kernel functions may result in the same classification accuracy. For example, when using testing set 2, the test accuracy of both Scheme 14 and Scheme 15 is 95.33%. Therefore, it can be concluded that, in most cases, Gaussian kernel function may be a good choice for combinations of kernel functions.

*6.1.4. Effect of Measurement Errors.* The "IEEE Standard for Synchrophasors for Power Systems" stipulates that PMUs with level 1 compliance should have a total vector error less than 1% [28]. For this reason, when using these kinds of PMUs to measure the bus voltages, the expected maximum error is 1%. For purpose of examining the predictive performance of the proposed method under measurement errors such as noise and other artifacts, a random error between 0 and 1% was added to all the original bus voltage measurements before extracting system-level classification features from them as inputs to the VBpMKL-based TSA model. The complete KB created in Section 6.1.1 was used for this test. In addition, same as in Scheme 16, Gaussian kernel functions are used for combinations of kernel functions.

Here, two schemes (Schemes 17-18) were designed and carried out to test the effects of measurement errors. In Scheme 17, the approach was evaluated without using these noisy data to train the predictive model. And then, the model was retrained using noisy data before using it to verify the predictive performance under measurement errors in Scheme 18. The results with noisy inputs, including Schemes 17 and 18, are illustrated in Table 6.

As observed in Table 6, the test accuracy in Scheme 17 was quite unsatisfying when without training with noisy inputs. At the same time, it shows that the proposal's performance of the presented approach in Scheme 18 has a slight decline compared to scheme 16 (the without noise case), but the test accuracy of the presented approach is, respectively, able to achieve 95.50% and 96.67% using testing sets 1 and 2, even when there are random measurement errors present in the input signals. Therefore, the conclusion can be drawn on the basis of the evidence that the proposed approach has good robustness and adaptability even under measurement errors.

*6.1.5. Test Results of Other Models.* To properly examine the proposal' performance, comparison tests between the present approach and other TSA models, such as DT, multilayer perception (MLP), and SVM, are carried out with the results summarized in Table 7. Here, all the abovementioned 23 features in Section 5 are employed as the inputs of these models.

The used models' parameters are set as follows. For DT, C4.5 algorithm is employed to construct the classification model for TSA; for MLP, the number of hidden layer nodes is set to 50 and the chosen training algorithm is backpropagation algorithm with the learning rate 0.8; for SVM, the chosen kernel function is the radial basis function, and the parameter selection approach is grid search with cross-validation techniques [16].

As shown in Table 7, the presented approach (Scheme 16) has better classification accuracy than all other TSA models, such as DT, MLP, and SVM. In addition, it can also be observed that, among the three other models, SVM has the best predictive performance. The reason for this is that SVM is a machine learning technique based on the statistical learning theory and the structural risk minimization principle. As a result, a conclusion which can be safely drawn from this is that the proposal is effective to predict the transient stability status of postfault systems.

*6.2. Case 2—The Power System of Hebei Province.* For the purpose of examining the applicability and effectiveness of the proposal to large-scale real-world power systems, the proposed approach is further tested on the system of Hebei province in China.

This tested system is made up of 83 generators, 650 major buses, and some series compensation devices and static var compensators. It is a highly interconnected large power system with a gross installed capacity of 28,260 MW, covering an area of 84,000 square kilometers.

*6.2.1. Generation of KB.* Same as in Case 1, plenty of simulations are executed to generate KB. Here, the used simulation calculation conditions (comprising the system model) of this system are the same as the one used in [9]. The transient stability criterion used here is consistent with that in Case 1. Finally, there are 2000 samples that are totally created through time-domain simulations. 1500 of the total samples are randomly chosen to constitute the training set and the remaining as the testing set.

*6.2.2. Prediction Results and Performance.* Similarly, the VBpMKL-based TSA model is trained, and then a contrast test using SVM is performed. In this case, the chosen combination of the feature subsets is $F_1$, $F_2$, and $F_3$; and the chosen combination of the kernel functions is the same combination as Scheme 16, $F_1(K_g)$, $F_2(K_p)$, and $F_3(K_p)$. The test results in this case are shown in Table 8.

Table 8: Test results in Case 2.

| TSA model | Feature subsets | Combinations of kernels | Test accuracy/% |
|---|---|---|---|
| VBpMKL | $F_1$, $F_2$, and $F_3$ | $F_1(K_g)$, $F_2(K_g)$, $F_3(K_g)$ | 97.20 |
| SVM | All features | — | 95.40 |

As is clearly seen in Table 8, the present approach is also applicable to large-scale actual systems. For one thing, the test accuracy of both TSA models in this case is slightly worse than that in Case 1; for another, the test accuracy of the proposal is obviously superior to that of SVM, same as in Case 1. These results demonstrate that with the increase of system scale and complexity, the chosen input features are required to increase consequently to more fully depict the stability characteristics of the disturbed system. Therefore, the presented method is a good choice to solve the issue of TSA for power systems.

## 7. Conclusions

PRTSA proves an effective way to determine the transient stability status of power systems, and wide-area state information plays a very important part in PRTSA. However, all traditional measurement systems are unable to provide the important measurements. The advent and matured applications of WAMS have made the availability of synchrophasors become reality, which brings new ideas and opportunities for implementing state-of-the-art WAPaC system. Consequently, in this paper we focus on taking full advantage of synchrophasors and propose a VBpMKL-based TSA approach through multifeature fusion.

Based on the simulation results on the test cases, the conclusions can be safely drawn as follows.

(1) The classification accuracy of the presented PRTSA model is significantly strengthened using multifeature fusion through combining feature spaces corresponding to each feature subset extracted from synchrophasors provided by WAMS.

(2) The predictive performance and generalization ability of the proposed method can be effectively improved by enhancing the completeness and representativeness of KB.

(3) The present algorithm might be employed as a trigger for a WAPaC system. Furthermore, it is possible to apply the methodology for constructing classification models to similar classification problems in the field of engineering.

## Nomenclature

TSA:    Transient stability assessment  
PRTSA:  Pattern recognition-based TSA

GPS:     Global positioning system
WAMS:   Wide-area measurement system
PMU:     Phasor measurement units
WAPaC:  Wide-area protection and control
SCADA:  Supervisory control and data acquisition
SIME:    Single machine equivalent
KB:      Knowledge base
MLP:     Multilayer perception
DT:      Decision tree
SVM:     Support vector machine.

## Conflict of Interests

The authors declare that there is no conflict of interests regarding the publication of this paper.

## Acknowledgments

This research is supported by the National Natural Science Foundation of China under Grant no. 51377016, the Doctor Scientific Research Foundation of Northeast Dianli University under Grant no. BSJXM-201407, and the Science and Technology Project of State Grid Corporation of China under Grant no. 2014GW-05 (the key technology research of a flexible ring network controller and its demonstration application).

10 Mathematical Problems in Engineering[26] S. K. Tso and X. P. Gu, "Feature selection by separability assessment of input spaces for transient stability classification based on neural networks," *International Journal of Electrical Power & Energy Systems*, vol. 26, no. 3, pp. 153–162, 2004.

[27] X. P. Gu, Y. Li, and J. H. Jia, "Feature selection for transient stability assessment based on kernelized fuzzy rough sets and memetic algorithm," *International Journal of Electrical Power & Energy Systems*, vol. 64, pp. 664–670, 2015.

[28] IEEE Standard, "IEEE Standard for synchrophasors for power systems," IEEE Std C37.118-2005 (Revision of IEEE Std. 1344-1995), 2006.

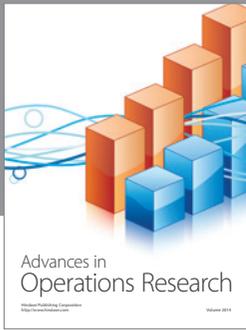
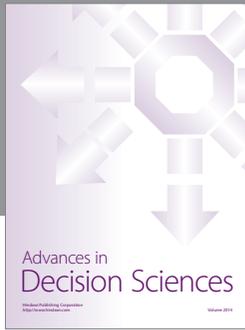
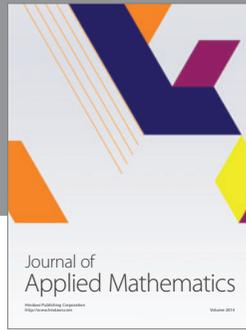
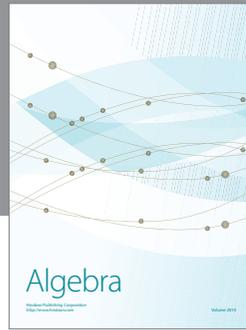
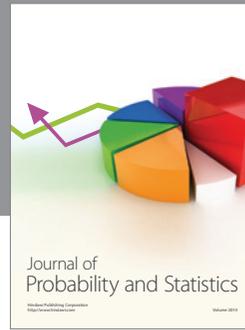
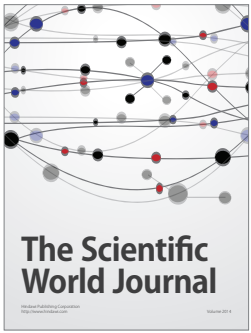
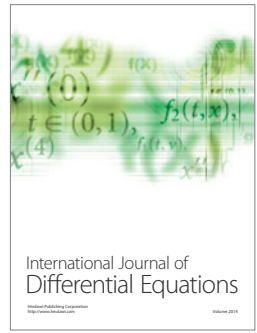
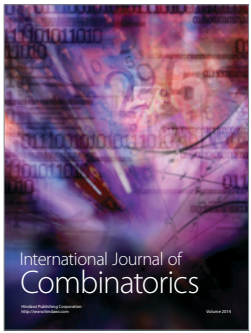
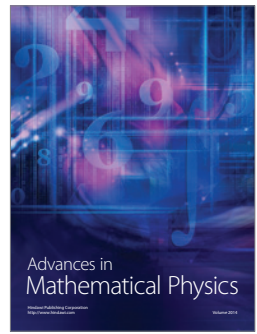
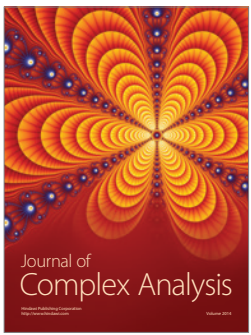
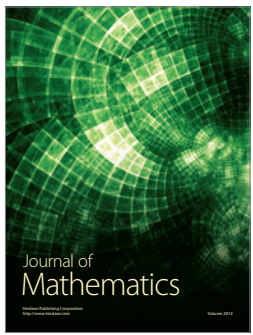
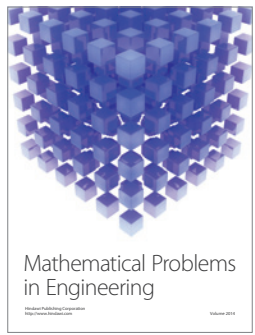
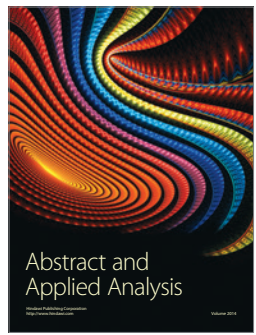
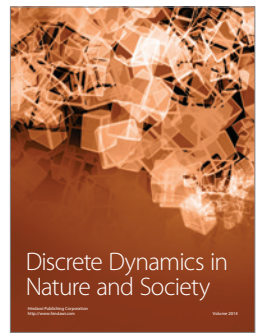
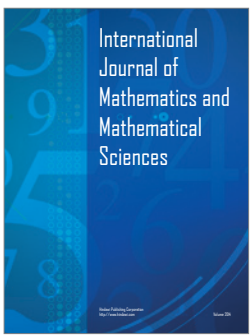
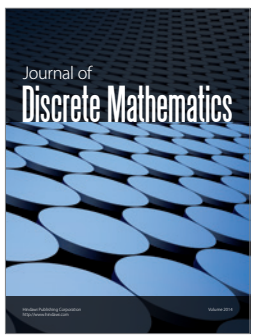
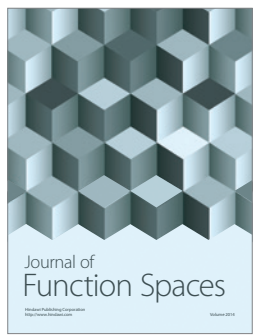
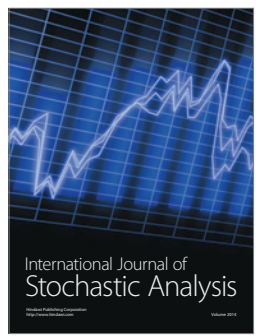
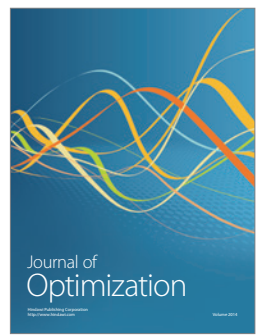

Submit your manuscripts at
http://www.hindawi.com